# Memory Efficient Quantum Circuit Simulator Based on Linked List Architecture

Abdel Samad, Wissam
Royal Institute of Technology
Stockholm, Sweden
wissamas@kth.se

Ghandour, Roy
Technical University of Munich
Munich, Germany
rng02@mytum.de

Hajj Chehade, Mohamad Nabil
University of California at Los Angeles
USA
nabilhc@ucla.edu

**Abstract**
In this paper, we will introduce the quantum circuit simulator we developed in C++ environment. We devise a novel method for efficient memory handling using linked list structures that enables us to simulate a quantum circuit of up to 20 qubits in a "reasonable" time. Our package can simulate the activity of any quantum circuit constructed by the user; it will also be used to understand the robustness of certain quantum algorithms (Simon, Shor).

## I. INTRODUCTION

In recent years interest in quantum computation has been steadily increasing. One reason for this is due to Shor's discovery of a polynomial time quantum algorithm for factoring integers, which is one of the strongest arguments in favor of the superiority of quantum computing models over classical ones. Since this discovery, many efforts have been made to find new, efficient quantum algorithms for classical problems and to develop quantum complexity theory. Actually in quantum systems, the computational space increases exponentially with the size of the system, which enables exponential parallelism. This parallelism could lead to exponentially faster quantum algorithms than possible classically.

**The Qubit**
In a classical computer, the logical state is determined by the value of register contents (voltage across a capacitor). The interpretation as (classical) bits is performed by comparing the measured value to a defined threshold, while the great number of particles guarantees that the uncertainty of the measurement is small enough to make errors practically impossible.

In a quantum computer, information is represented as the common quantum state of many subsystems. Each subsystem is described by a combination of two "pure" states interpreted as |0> and |1> (quantum bit, qubit). Physically, this can be realized by the spin of a particle, the polarization of a photon or by the ground state and an excited state of an ion.

For a single qubit, this state can be described by the complex amplitudes $a$ and $b$ of each of the two states ($a|0> + b|1>$) where $|a|^2$ and $|b|^2$ are the respective probabilities for the qubit to be in state |0> and state |1> with the condition $|a|^2 + |b|^2 = 1$. It is obvious, that this interpretation stands in contradiction to classic Boolean logic, where intermediate states between 0 and 1 are not possible. Qubits are represented by the basis vectors. So |0> is represented by the vector (1 0), |1> by (0 1), |00> by (1 0 0 0) and so on. Also the vector $(1/\sqrt{2}, 1/\sqrt{2})$ has a probability of 0.5 to be in state |0> and 0.5 to be in |1>. Quantum computation is based on two quantum phenomena: quantum interference and quantum entanglement. Entanglement allows one to encode data into

non-trivial multi-particle superposition of some preselected basis states, and quantum interference, which is a dynamical process, allows one to evolve initial quantum states (inputs) into final states (outputs) modifying intermediate multi-particle superposition in some prescribed way. Multi-particle quantum interference, unlike single particle interference, does not have any classical analogue and can be viewed as an inherently quantum process. While a classical *N*-bit register can hold any *one* integer between 0 and $2^N-1$, an *N*-quantum bit (q-bit) register can simultaneously hold *all* of these $2^N$ integers in the form of a quantum *superposition* (combination) of states. However, there is a catch—the result of the computation is also in the form of a superposition and any attempt to read out (measure) the result causes the superposition to collapse into just one of the $2^N$ integer states. The measurement of a particular quantum state will not deterministically collapse to the same state. The collapse to different states is probabilistic. Therefore, to get more information on the quantum state being measured, one must freshly prepare the superposition many times and perform the measurement each time. The outcomes of each of these "trials" will allow one to build a detailed understanding of the likelihood of the quantum state to collapse to each of the different possible states.

## II. THE QUANTUM CIRCUIT

To understand the underlying structure of quantum circuit, we first need to know how to represent qubits and their operators (gates).

An *N-qubit* register is represented by a $2^N$ complex vector $(a_1, a_2, ..., a_n)$ called the qubits' state vector where $|a_i|^2$ is the probability of finding the system of these N qubits in a given state $\Psi_i$.

A quantum gate on *k* qubits is defined to be a $2^k$ x $2^k$ unitary matrix[1] *U*. This matrix U is equivalent to a quantum operator in theoretical quantum physics. When qubits are passed through a gate, we multiply the qubits' state vector by the gate's matrix U to obtain the qubits' new state.

### A. Basic Gates

Quantum gates are the building blocks of a quantum circuit. Next we present the basic gates we implemented and used in our simulations

### 1. NOT gate

In conventional logic, a NOT gate inverts the value of a bit from 1 to 0 or from 0 to 1. Now in quantum logic, the NOT gate inverts the probabilities of the qubits' state vector. The one qubit NOT gate can be represented by the following 2x2 matrix:

$$\begin{pmatrix} 0 & 1 \\ 1 & 0 \end{pmatrix}$$

Therefore when a qubit $|q\rangle= (q_1,q_1)$ is inputted to a NOT gate, it will be transformed to $|q'\rangle=(q'_1,q'_1)= (q_2,q_1)$.

The N-qubit NOT gate's matrix is the tensor product of N one-qubit NOT gates:

$$\begin{pmatrix} 0 & 1 \\ 1 & 0 \end{pmatrix} \otimes \begin{pmatrix} 0 & 1 \\ 1 & 0 \end{pmatrix} \otimes \cdots \otimes \begin{pmatrix} 0 & 1 \\ 1 & 0 \end{pmatrix}$$

which is equivalent to:

$$\begin{pmatrix} 0 & 0 & \cdots & 0 & 1 \\ 0 & \cdots & \cdots & 1 & 0 \\ \vdots & \vdots & \iddots & 0 & \vdots \\ \vdots & 1 & \cdots & \cdots & 0 \\ 1 & 0 & \cdots & \cdots & 0 \end{pmatrix}$$

## 2. Hadomard Gate

A Hadomard gate is one of the most important gates. The one-qubit Hadomard gate's matrix is:

$$H = \frac{1}{\sqrt{2}} \begin{pmatrix} 1 & 1 \\ 1 & -1 \end{pmatrix}$$

As for the N-qubit NOT gate, the N-qubit Hadomard gate's matrix is the tensor product of N one-qubit

Hadomard gates:

$$H \otimes H \otimes \ldots \otimes H$$

## 3. C-not Gate

The *controlled-NOT (C-NOT)* gate is a 2-qubit quantum gate. The second qubit *(target bit)* is negated only if the first qubit *(control bit)* is set. The C-not gate can be modeled as follows:

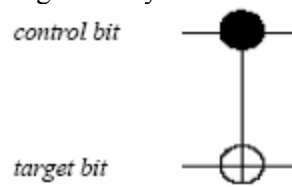

and its matrix is:

$$\begin{pmatrix} 1 & 0 & 0 & 0 \\ 0 & 1 & 0 & 0 \\ 0 & 0 & 0 & 1 \\ 0 & 0 & 1 & 0 \end{pmatrix}$$

## 4. Quantum Fourier Transform *(QFT)*

The quantum Fourier transform *(QFT)* is the principal algorithmic tool underlying most efficient quantum algorithms. The Fourier transform is a linear transform, and can be represented by an *n* x *n* matrix *M* where

$$M_{x,y} = \frac{1}{\sqrt{n}} \omega^{xy}$$

and $\omega = e^{\frac{2\Pi i}{n}}$

On the quantum scale, the Fourier transform (QFT) is the transformation on states:

$$|x\rangle \rightarrow \frac{1}{\sqrt{n}} \sum_{y=0}^{N-1} \omega^{xy} |y\rangle$$

where $x$ and $y$ are basis vectors of $n = \log N$ bits, interpreted as a number in binary.

### B. Circuit Architecture

A quantum circuit is a succession of stages, where each stage is a parallel combination of gates. The input of one stage is the output of the previous one. In the implementation, we conserved this structure in the memory: the simulator interprets each stage as being a linked list (Gate LList) whose nodes hold the matrices of the gates forming the stage. The succession of stages is implemented by means of another linked list (Circuit-LList) where each node points to a Gate-LList. The underlying lattice of the circuit will look as the following figure:

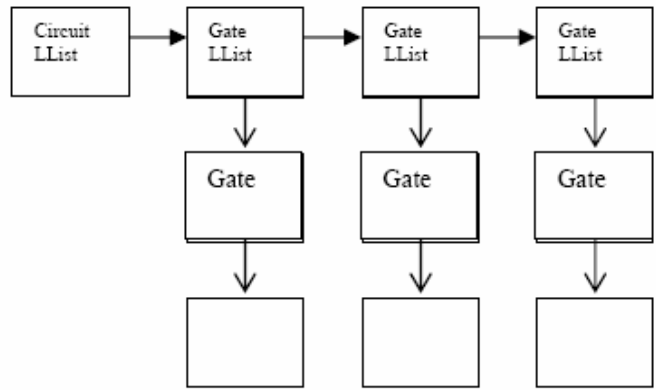

### Implementations

The job of the simulator is to get an output based on an input and a circuit. The input and the output (prior to the measurement) are qubits vectors. As a start, we implemented a rather naïve method which was a necessary step to explore a more efficient method.

### The naïve implementation

The straightforward way to calculate the output vector is to multiply the input vector by the representative matrix of the whole circuit. Consecutively, we tensor the gate matrices in the gate linked lists starting top to bottom and store the results (stage matrix) in the header nodes. Then these matrices are multiplied together from left to right resulting in the overall matrix of the circuit. At this stage, we can get the output of the circuit by just multiplying the input register and the overall matrix.

In this implementation, we notice that time and space complexities grow exponentially: a circuit composed of many stages; with each stage having $L$ gates of size $A \times A$ an amount of order $A^{2L}$ of memory is needed. If $A=8$, 1 Gigabytes of memory is consumed.

### Efficient implementation

The second approach tries to handle this issue. In fact, the first entry in the output vector of one stage is the scalar product of the input vector with the tensor of the first columns of the $L$ matrices forming the stage.

Generically, the second entry is equal to the product of the input vector with the tensor of the first columns of the first $L-1$ matrices and the second column of the last matrix. We implemented this method reducing the memory usage to an order of $A^L$. So now instead of saving the whole circuit matrix in memory, we are only saving one column of that matrix and using it to find its corresponding output entry. Then this column in memory is destroyed and the next column of the circuit matrix is saved and so on. Thus, this new approach reduced memory usage by the order of $N$, where $N$ is the size of the circuit matrix. The table below shows the results obtained after simulating test quantum circuits of different number of inputs. It is clear that the memory saving is large. Notice that in the naïve case, a maximum of 12 qubits were possible to simulate, above which the computer crashed. The simulations were performed on a PIV 2.26 GHz PC with 512 Mbytes RAM.

| Qubits | Time | | Memory Usage | |
|---|---|---|---|---|
| | Naïve | Efficient | Naïve | Efficient |
| 6 | 120msec | 120msec | 1 MB | 62KB |
| 10 | 4 sec | 2.6 sec | 171 MB | 136KB |
| 11 | 33.4 sec | 10.3 sec | 898 MB | 262KB |
| 12 | 143 sec | 55 sec | 1.71 GB | 0.34MB |
| 13 | Crash | 155 sec | Crash | 0.71MB |
| 14 | Crash | 626 sec | Crash | 1.3MB |
| 15 | Crash | 37 min | Crash | 1.8MB |
| 16 | Crash | 2.73 hours | Crash | 3MB |
| 17 | Crash | 10.92 hours | Crash | 5MB |
| 18 | Crash | 44.41 hours | Crash | 10MB |
| 19 | Crash | 176.1 hours | Crash | 15MB |
| 20 | Crash | 307.2 hours | Crash | 34MB |
| 21 | Crash | 17.8 days | Crash | 53MB |
| 22 | Crash | Undetermined | Crash | 124MB |
| 23 | Crash | Undetermined | Crash | 506MB |
| 24 | Crash | Undetermined | Crash | 768MB |
| 25 | Crash | Undetermined | Crash | 1.23GB |

The final step is to apply a quantum measurement to the output register. Postulates of quantum mechanics posit that measurement is probabilistic and collapses the superposition to a single bit-string — the larger the amplitude of a bit-string, the more likely the bit-string is to be observed. The procedure we used for measurement was simple and proved to be quite random. We use a random number generator that generates a value between zero and one. We then start adding the probabilities of each state until the sum becomes greater than or equal to the random number generated. The final state whose probability was added is the state that is considered the output of the measurement. We tested our simulator on two of the most important and basic algorithms, Shor's algorithm for prime factorization and Simon's for determining if a function is 1-1 or 2-1 in polynomial time.

## III. SHOR'S ALGORITHM

**A. Motivation**
While it is quite easy to find and multiply prime numbers, the reverse process is believed to be practically impossible. Until now, no efficient algorithm has been found for factoring a large number on a classical computer. (An algorithm is called efficient if the number of elementary operations is asymptotically polynomial in the length of its input). The best-known algorithm grows exponentially with the input size. This fact is essential in public key cryptosystems where the key for encoding is public and the decoding key remains secret. Another interesting problem where no efficient classical algorithm has been found is finding the order of an element. Given integers $x$ and $n$, the order $r$ is such that $x^r \equiv 1 \pmod{n}$.

**B. Algorithm**

In 1994, P.W. Shor proposed an efficient algorithm for finding the order of an element on quantum computers. The factoring of a number can now be done as follows:

*Input*: n is odd and none prime integer to be factorized
*Step 1*: choose a random number x where $0 \leq x \leq n$
*Step 2*: use Shor's algorithm to find the order r of x (mod n)
*Output*: if r is odd or $x^{r/2} \equiv -1 \pmod{n}$ then failure
Else return a factor = $\gcd(x^{r/2} - 1, n)$

*Proof*:
First we notice that choosing a random number between 0 and n (in step one), doing the modular exponentiation and finding the gcd (in step three) can be done in polynomial time. We will also prove next that step two is also done in polynomial time. Thus the whole algorithm is done in polynomial time.
Next, we know that if n is a composite number and $x^2 \equiv y^2 \pmod{n}$ and $x \neq \pm y \pmod{n}$, then x-y is a factor of n. So if we find the order of an integer x (mod n), then $x^r \equiv 1^2 \pmod{n}$, then $x^{r/2} \equiv 1 \pmod{n}$, and the factor will be $\gcd(x^{r/2} -1, n)$.

Steps of Shor's algorithm:
On input a prime p and a generator g and some number x, we need to find r so that $g^r \equiv x \pmod{p}$
1) Start with the state $|0\rangle \otimes |0\rangle \otimes |0\rangle$. (Here, each register can store numbers from 0 to p-2)
2) Apply QFTp-1 to the first two registers, we reach the state:

$$(1/p-1) \sum_{a=0}^{p-2} \sum_{b=0}^{p-2} |a,b\rangle \otimes |0\rangle$$

3) Replace the final register with the value $g^a x^{-b} \pmod{p}$. Thus we reach the state:

$$(1/p-1) \sum_{a=0}^{p-2} \sum_{b=0}^{p-2} |a,b\rangle \otimes |g^a x^{-b} \pmod{p}\rangle$$

4) Apply QFTp-1 again to the first two registers, we reach the state:

$$(1/p-1)^2 \sum_{a=0}^{p-2} \sum_{b=0}^{p-2} \sum_{c=0}^{p-2} \sum_{d=0}^{p-2} w^{ac+bd} |c,d\rangle \otimes |g^a x^{-b} \pmod{p}\rangle$$

5) Measure the entire register and check if $\gcd(c,p-1)=1$. If yes, find r by solving $c*r - d \equiv 0 \pmod{p-1}$; if no, run the circuit until the condition is satisfied.

## IV. SIMON'S ALGORITHM

*Claim*: Given any function $f : \{0,1\}^n \to \{0,1\}^n$, we can build a quantum circuit $Q_f : \{0,1\}^n$ x $\{0,1\}^n \to \{0,1\}^n$ x $\{0,1\}^n$ which computes $(x,y) \to (x, y \oplus f(x))^2$.

*Theorem*: Given the above function $f$ and a promise that $f$ is either 1-to-1 or 2-to-1

(ie $\exists t : \forall x, f(x) = f(x \oplus t)$ ), we can determine which case is true in quantum polynomial time. In the 2-to-1 case, we can find t.

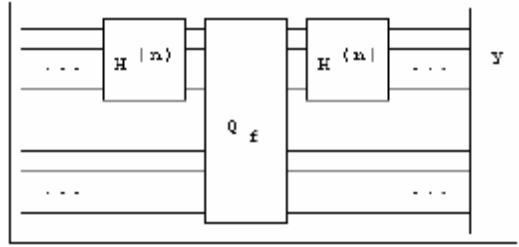

1) We start with an input of 2n qubits all initialized to zero:
2) After the first Hadomard gate, we arrive at the state:

$$\sum_{x=0}^{2^n-1} 2^{-n/2} |x> \otimes |00...0>$$

3) After $Q_f$ we arrive at:

$$\sum_{x=0}^{2^n-1} 2^{-n/2} |x> \otimes |f(x)>$$

4) Apply the second Hadomard gate:

$$\sum_{x=0}^{2^n-1} \sum_{y=0}^{2^n-1} 2^{-n} (-1)^{x \bullet y} |y> \otimes |f(x)>$$

5) Repeat this routine 2n times and measure the top n qubits each time. In the 1-to-1 case we will have a uniform distribution of all the states each with a coefficient of $\pm 2^{-n}$. In the 2-to-1 case, the states are identical and thus have the same coefficient:

$$\alpha = \frac{(-1)^{x \bullet y} + (-1)^{(x \oplus t) \bullet y}}{2^n}.$$

If $t \bullet y \equiv 1 \pmod 2$, then $\alpha = 0$ and the state does not exist. Otherwise the state will exist. Thus, after 2n measurements, we will obtain a system of equations of the form $y_i \bullet t \equiv 0 \pmod 2$. If the number of distinct equations is greater or equal to n; i.e. $i \geq n$, then the system can be solved by Gaussian elimination to yield a unique vector $t$. Otherwise, the system can not be solved uniquely. We tested Simon's algorithm for values of n = 3,4, and 5 qubits. To increase the probability of finding $t$, we repeated step 5 of the routine for *3n* times instead of *2n*. In the case of *n=3*, *t* was found 90% of the times, for n=4 and 5, t was found 99% of the times.

## V. CONCLUSION

An efficient simulator was built. We were able to simulate up to 20 qubits in a reasonable time. Moreover we tested Simon's and Shor's algorithms and successful results were obtained.

## VI. ACKNOWLEDGMENTS

The authors would like to express their great thanks to Prof. Louay Bazzi for his helpful discussions and clarifications throughout this project.